\documentclass[structabstract]{aa}  
\usepackage{graphicx}
\usepackage{txfonts}
\usepackage{natbib}
\bibpunct{(}{)}{;}{a}{}{,} 

\begin{document}
 \title{\textit{Herschel}\thanks{The \textit{Herschel} data described in this paper
have been obtained in the open time project {\tt OT1\_tpreibis\_1} (PI: T.~Preibisch). \textit{Herschel}
is an ESA space observatory with science instruments provided by European-led Principal Investigator
consortia and with important participation from NASA.}
 far-infrared observations of the Carina Nebula complex
}
\subtitle{ {\rm I:} Introduction and global cloud structure}

   \author{T.~Preibisch\inst{1} \and V.~Roccatagliata\inst{1} \and B.~Gaczkowski\inst{1}
           \and T.~Ratzka\inst{1} 
          }

   \institute{Universit\"ats-Sternwarte M\"unchen, Ludwig-Maximilians-Universit\"at,
          Scheinerstr.~1, 81679 M\"unchen, Germany \email{preibisch@usm.uni-muenchen.de} }

\date{Received 19 January 2012; accepted 13 March 2012}

 
  \abstract
   {The Carina Nebula represents one of the most massive star forming regions known in 
   our Galaxy and displays a high level of feedback from the large number
   of very massive stars. While the stellar content is now well known from 
  recent deep X-ray and near-infrared surveys, the properties of the clouds
   remained rather poorly studied until today.
   }
   {By mapping the Carina Nebula complex in the far-infrared, we aim at a comprehensive and 
  detailed characterization of the dust and gas clouds in the complex.
   }
   {We used SPIRE and PACS onboard of \textit{Herschel}
   to map the full spatial extent ($\approx 5.3$~square-degrees) 
 of the clouds in the Carina Nebula complex at wavelengths between
  $70\,\mu$m and $500\,\mu$m. We use here the $70\,\mu$m and $160\,\mu$m
   far-infrared maps to determine color temperatures and 
column densities, and to investigate the global properties of the
 gas and dust clouds in the complex.
   }
   {Our \textit{Herschel} maps show the far-infrared morphology of the clouds at unprecedented
   high angular resolution.
  The clouds show a very complex and filamentary structure that is dominated
  by the radiation and wind feedback from the massive stars. In most locations, the
column density of the clouds is $N_{\rm H} \la 2 \times 10^{22}\,\rm cm^{-2}$ (corresponding to visual extinctions
 of $A_V \la  10$~mag); denser cloud structures are restricted to the massive cloud west of Tr~14 and
 the innermost parts of large pillars.
Our temperature map shows a clear large scale gradient 
from  $\approx 35 - 40$~K in the
central region to $\la 20$~K at the periphery and in the densest parts of individual pillars.
The total mass of the clouds seen by \textit{Herschel} in the central (1 degree radius) region
is $\approx 656\,000\,M_\odot$.
We also derive the global spectral energy distribution in  the mid-infrared to mm wavelength range.
A simple radiative transfer model suggests that
the total mass of all the gas (including a warmer component that is not well traced by 
\textit{Herschel}) in the central 1 degree radius region is $\le 890\,000\,M_\odot$.
   }
   {
Despite the strong feedback from numerous massive stars and the corresponding cloud dispersal
processes that are going on since several 
million years, there are still several $10\,000\,M_\odot$ of cool cloud material present 
at column-densities sufficient for further star formation.
Comparison of our total gas mass estimates to molecular cloud masses derived from CO line mapping 
suggests that as much as about 75\% of all the gas is in atomic rather than molecular form.
}


   \maketitle
%

\section{Introduction}

Most stars form in large clusters or associations, containing
(at least) several thousand stars \citep[e.g.,][]{Briceno07}, and it
has recently become clear that also our sun formed as part of
a large star forming complex \citep[e.g.,][]{Adams10}.
In contrast to low-mass star forming regions like Taurus, where the
individual young stellar objects (YSOs) form more or less in isolation
and any interaction is minimal, large clusters and associations
contain high-mass ($M \ge 20\,M_\odot$) stars.
These hot and luminous O-type stars 
profoundly influence their environments \citep[see, e.g.][]{Freyer03} by creating 
HII regions, generating wind-blown bubbles, and exploding as 
supernovae. This feedback can disperse the natal molecular clouds and thus
halt further star formation. However, advancing ionization fronts 
and expanding superbubbles can also compress nearby clouds and 
thereby trigger the formation of new generations of stars
\citep[e.g.,][]{Reach04,Cannon05,Oey05,PZ07,Deharveng09,Zavagno10,Brand11}.
The interaction of the massive stars with the surrounding clouds and the 
balance between cloud destruction and triggered star formation
determine the characteristics and the final outcome of the
star formation process
\citep[see][for numerical studies]{Dale07,Dale08,Bate09,Gritschneder10}.

At a distance of 2.3~kpc, the Carina Nebula Complex 
\citep[CNC, hereafter; see][for a recent review]{SB08} 
is the nearest star forming
region with a large population of very massive stars 
\citep[at least 70 O-type and WR stars; see][]{Smith06}.
Among these are several
of the most massive  ($M \ga 100\,M_\odot$) and luminous stars 
known in our Galaxy, e.g.,~the
famous Luminous Blue Variable $\eta$\,Car, 
the O2 If* star
HD~93129Aa, several O3 main sequence stars, and four Wolf-Rayet  stars.
Most of the massive stars reside in several
loose clusters near the center of the complex,
including Tr~14, 15, and 16,
which have ages ranging from $<$1 to $\sim 8$~Myr
\citep[see][]{Dias02,HAWKI-survey}.
The CNC contains large amounts of gas and dust clouds,
with an estimated total mass in the range
$350\,000 - 670\,000\,M_\odot$ based on CO radio maps \citep[e.g.][]{Grabelsky88,Yonekura05,SB07}.
The very strong radiative and wind feedback from the
massive stars has already dispersed much 
of the original cloud mass in the central region, and drives a large expanding superbubble,
extending over $\sim 80$~pc (corresponding to $\sim 2\degr$).

During the last few years, 
several surveys have provided a wealth of new information
on the stellar populations in the CNC.
The \textit{HST} survey  of the central area \citep{Smith10a}
and a \textit{Spitzer} survey of the southern parts of the CNC \citep{Smith10b}
showed that the ionizing radiation from the massive stars is currently
triggering the formation of a new generation of stars in the remaining dense
clouds, in particular in the so-called ``Southern Pillars'' region.

We have performed a very deep near-infrared
survey of the central $\sim 1300$ square-arcminute area of the CNC
with the near-infrared camera HAWK-I at the ESO 8m-VLT \citep{HAWKI-survey} which
is deep enough to
reveal {\em all} young stars
in the region with $A_V \leq  35$~mag.

The {\it Chandra} Carina Complex Project \citep[see][for an overview]{CCCP-intro} has
mapped the CNC with a mosaic of 22  individual
ACIS-I pointings, using a total observing time of
1.34 Mega\-seconds (15.5 days) and covering an area
of about 1.4 square-degrees.
With the detection of more than
11\,000 X-ray emitting young stars, these \textit{Chandra} data efficiently
eliminate the strong field-star confusion problems that plague visual and
IR samples and provide, for the first time, a large
sample of the young stellar population (down to $\sim 0.5\,M_\odot$)
in the area. A detailed statistical analysis of the X-ray, optical, and
infrared properties of the detected sources showed that
10\,714 of the X-ray sources
are very likely young stars in the CNC \citep{CCCP-classification}.
The analysis of the spatial distribution of the X-ray detected young stars showed
that half of the young stellar population resides in one of about 30 clusters or stellar
groups, while the other half constitutes a widely dispersed population 
\citep{CCCP-Clusters}. The combination of the X-ray data with near- and mid-infrared photometry
provided new information about the properties of the stellar populations
in the entire complex \citep{Povich11,CCCP-HAWKI} and the individual clusters Tr~14 and
Tr~15 \citep{CCCP-Tr14,CCCP-Tr15}. These studies showed that the
X-ray detected young stellar populations
have ages ranging from $< 1$~Myr up to $\approx 8$~Myr,
and support the scenario of ongoing, triggered star formation in the CNC.

While these new data sets have strongly boosted the amount of available
information about the stellar populations, the knowledge about the
clouds in the complex is still much more limited.
 While the IRAS maps and several existing radio maps of CO and
other molecular lines provide basic information
on the large-scale morphology of the clouds, only very small
areas at the center of the complex have been observed with sub-arcminute
angular resolution at far-infrared (FIR) and (sub)-mm wavelengths so far
\citep{Brooks05,Gomez10}.
This lack of sensitive FIR and (sub)-mm data with sufficient spatial
resolution 
strongly hampers studies of the cloud structure, and
is a very serious obstacle for
investigations of global cloud properties and the
interaction between massive stars and clouds in the CNC.

As a first step to improve this situation,
we have recently used LABOCA at the APEX telescope to obtain a sensitive 
wide-field
($1.25\hbox{$^\circ$} \times 1.25\hbox{$^\circ$}$) sub-mm
map of the CNC at a wavelength of $870\,\mu$m with
$18''$ angular resolution ($=$ 0.2~pc at the distance of the CNC), which
 provides the
first large-scale survey of the dusty clouds in this region
\citep{CNC-Laboca}.
These data showed that there are (at least) 
$\sim 60\,000\,M_\odot$ of dust and gas in dense, compact clouds;
this represents a large potential for further star formation.
While this LABOCA map provided important information about the
global properties of the dense clouds in the complex, the information
that can be retrieved about the total cloud phase is limited by 
two factors:
First, the LABOCA map is only sensitive to the dense, localized
clouds. As a consequence of the removal of correlated noise
in the data reduction, structures with angular sizes larger $\approx 2.5'$
are only partly recovered. Therefore, the more diffuse emission from
less dense gas is missing from the map.
The second  limitation is that a 
single-wavelength band map does not allow the determination of 
cloud temperatures. This is critical, because information on the
cloud temperature is required to compute reliable
cloud masses from the observed fluxes. The often used approach
of simply assuming a uniform ``typical'' cloud
temperature (e.g.,~15~K) is very likely not appropriate in the case of the CNC, where
some clouds are very strongly
irradiated (and thus heated) by numerous nearby massive stars,
while other clouds (especially at the periphery of the CNC)
experience orders of magnitudes lower levels
of irradiation (and correspondingly less heating from outside).
FIR maps at several different wavelengths can provide
crucial information about the cloud temperatures 
and thus allow much more reliable mass and column-density estimates
than based on single-wavelength data.

In the present paper we will be discussing observations of the CNC performed with the 
ESA \textit{Herschel} Space Observatory \citep{Pilbratt10}, 
in particular employing Herschel's large telescope and powerful science payload 
to do photometry using the PACS \citep{Poglitsch10} and SPIRE \citep{Griffin10} instruments.
We will present an overview of the project, determine color temperature and column density
maps and investigate the statistics of these quantities in selected regions of our maps.
We also investigate global properties of the CNC clouds derived from our results. 

\section{\textit{Herschel} far-infrared mapping of the Carina Nebula}

In the \textit{Herschel} open time project {\tt OT1\_tpreibis\_1}
we  used PACS and SPIRE to map
the entire extent of the CNC in five bands between 
 $70\,\mu$m and $500\,\mu$m.
The observation was performed on 26 December 2010 in the SPIRE PACS Parallel Mode
and used 6.9 hours observing time. The J2000 coordinates of the aimpoint are 
RA = 10h44m03s, Dec = -59d30m00s.
The fast scan speed (60 arcsec per second) was used to 
map a  $\approx 2.3^\circ \times 2.3^\circ$ area\footnote{Note that 
due to the fixed $21'$ separation of the SPIRE and PACS focal plane
footprints on the sky, the areas observed by each individual instrument 
are somewhat larger and show some offset with respect to each other;
the mentioned $\approx 2.3^\circ \times 2.3^\circ$ area is the
region that is covered by both SPIRE \textit{and} PACS.}, which 
corresponds to a physical region of $\approx 92\,\rm pc \times 92\, pc$ at the distance
of the CNC and
covers the {\em full spatial extent} of the CNC.
The mapped area also includes the HII region Gum~31 \citep[= Ced~108 = NGC~2599; see][]{Cederblad46,Gum55}
 around the young stellar cluster NGC~3324
at the north-western edge of the Carina Nebula.

The observation was done with two scan maps of the same area, one with nominal
scan direction, and the other with orthogonal scan direction,
in order to remove more efficiently the stripping effects due to the
$1/f$ noise and to get better coverage redundancy.

We used the  $70\,\mu$m PACS  channel as the blue band, in order
to get the widest possible wavelength range.
The wavelength bands of the five maps resulting from our
observations are $[60 - 85]\,\mu$m, $[130 - 210]\,\mu$m, $[210 - 290]\,\mu$m,
$[300 - 400]\,\mu$m, and $[420 - 610]\,\mu$m, and will be denoted as
the $70\,\mu$m, $160\,\mu$m,  $250\,\mu$m,  $350\,\mu$m, and  $500\,\mu$m band
in the following text.

The data reduction was performed with the \textit{HIPE} v7.0 \citep{hipe}  and
\textit{SCANAMORPHOS} v10.0 \citep{Roussel12} 
software packages\footnote{see {\tt http://www2.iap.fr/users/roussel/herschel/}}.
From level 0.5 to 1 the PACS data was reduced using the L1\_scanMapMadMap script in the photometry pipeline 
in HIPE with the version 32 calibration tree.
The level 2 maps were produced with \textit{SCANAMORPHOS} with standard options for parallel mode 
observations and the {\tt galactic} option to preserve brightness gradients over the field. 
The pixel sizes of the two PACS maps were chosen as $3.2\arcsec$ (for the $70\,\mu$m map)
 and $4.5\arcsec$ (for the $160\,\mu$m map), as suggested in \citet{hi-gal-reduction}.

The level 0 SPIRE data were reduced with an adapted version of the HIPE script 
rosette\_obsid1\&2\_script\_level1 included in the \textit{SCANAMORPHOS}  package. 
The final maps were produced by \textit{SCANAMORPHOS} with standard options for parallel mode observations 
and the {\tt galactic} option as for the PACS data. The pixel sizes were chosen 
as $6\arcsec$, $8\arcsec$ and $11.5\arcsec$, for the 250, 350, and 500 $\mu$m band, respectively.

The quality of the final maps turned out to be very good.
In order to characterize the point-spread-function (PSF) in the maps,
we determined the full-width at half-maximum (FWHM) for a number of isolated
point-like sources and found values of $\approx 10''$ for the $70\,\mu$m map,
$\approx 15''$ for the $160\,\mu$m map, $\approx 20''$ for the $250\,\mu$m map,
$\approx 26''$ for the $350\,\mu$m map, and $\approx 36''$ for the $500\,\mu$m map.
These values show that the image quality of the PACS maps is slightly 
worse than the telescope diffraction limit
(as expected for observations performed in fast scan mode),
but the SPIRE maps are close to the ideal quality.
With their unprecedented angular resolution of $\sim 10''-36''$,
corresponding to physical scales of $\sim 0.1-0.4 $~pc,
these \textit{Herschel} maps represent the first detailed, deep,
and spatially complete FIR maps of the full CNC.

The levels of background cloud fluctuations,
determined as the standard deviation of the pixel value in apparently
empty regions in the maps, are found to be $\approx 2.3$~mJy/px at $70\,\mu$m, 
$\approx 5.0$~mJy/px at $160\,\mu$m, 
$\approx 84$~mJy/beam at $250\,\mu$m,
$\approx 83$~mJy/beam at $350\,\mu$m, and $\approx 73$~mJy/beam at $500\,\mu$m.

A potential problem in the calibration of the \textit{Herschel} maps is the
possible presence of zero-level offsets \citep[see][]{Bernard10}. 
No correction for these offsets is currently implemented
in the \textit{Herschel} data reduction tools.
There are several reasons why we can expect the uncertainties related to offsets 
to be only moderate for our specific data set and the results obtained in  
this paper.
One reason is that 
by the use of the the {\tt galactic} option in \textit{SCANAMORPHOS} the baseline
subtraction is modified with the aim to preserve large-scale structures as well as possible.
Another important aspect is that our maps cover the entire
spatial extent of the clouds associated with the Carina Nebula. 
The periphery of our maps shows low-intensity galactic background\footnote{Since
the Carina Nebula is close to the galactic
plane ($b \sim -0.6\degr$) and near the tangent point of a spiral arm,
the background consists of numerous distant cloud complexes.}
 emission that is not related
to the Carina Nebula and thus not relevant for the aims of our study.
Therefore, the emission from the relevant clouds associated to the Carina Nebula
can be well separated from the galactic background in our maps.
We also compared our \textit{Herschel} $70\,\mu$m map to the
IRAS $60\,\mu$m map and found very good agreement (within about 1\%) of the
fluxes integrated over large regions.
Finally, we note that in our temperature and column density determination described
in the next section even a
20\% error in the intensity causes just a $\sim 5\%$ change in the derived
color temperatures and a $\sim 12\%$ change in the derived cloud column densities.

As soon as the \textit{Planck} data get available, a more detailed analysis 
of the possible effects of 
offset corrections will be performed.
%
%

   \begin{figure*}
   \centering
   \includegraphics{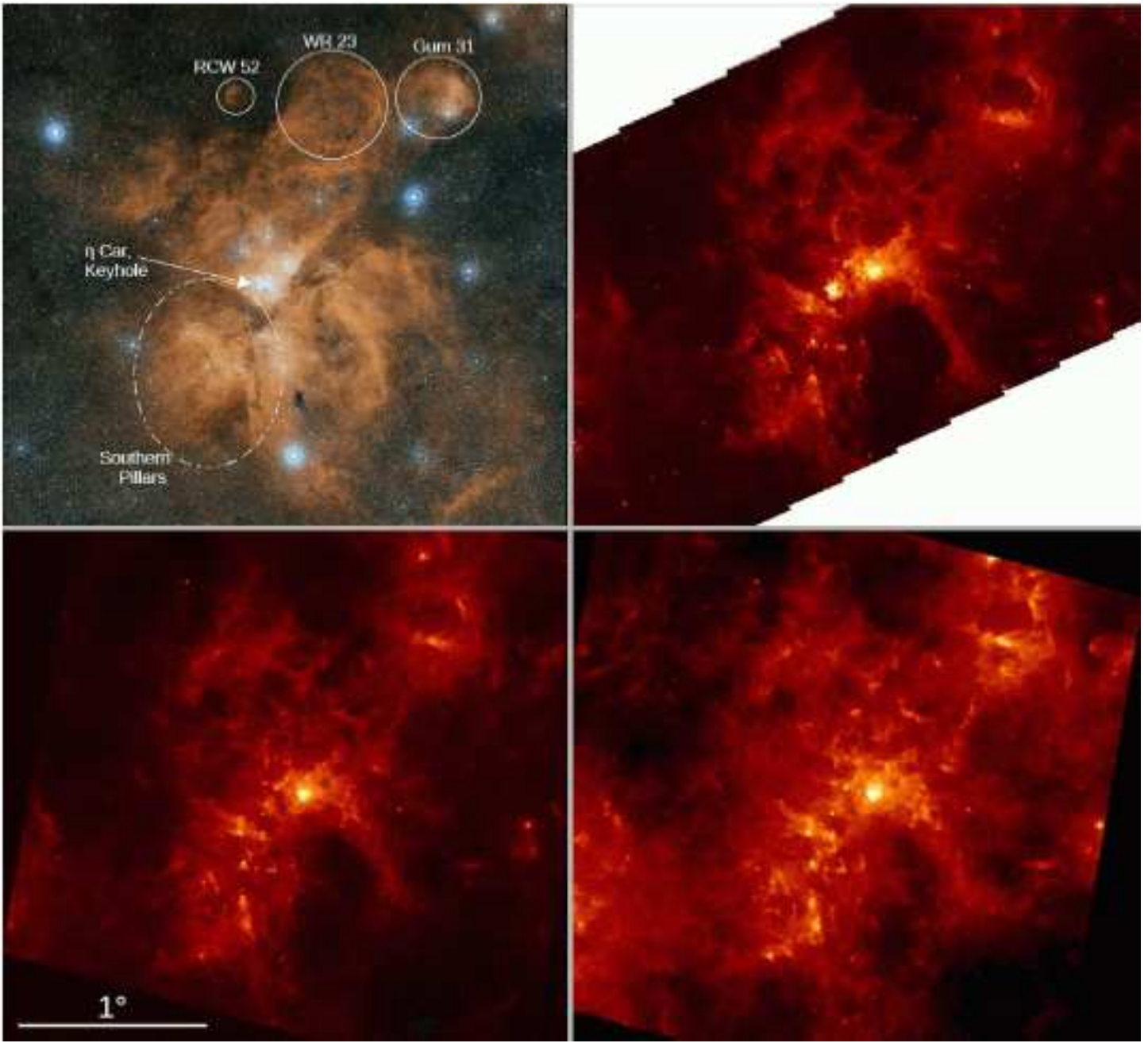}
   \caption{Optical, near-, and far-infrared view of the Carina Nebula.
All four images show the same field of view of $3\degr \times 3\degr$.
North is up and east to the left.
\textbf{Upper left:} Optical image obtained from the ESO Photo Release 1145
(http://www.eso.org/public/images/eso0905b/; ESO/Digitized Sky Survey 2, Davide De Martin).
\textbf{Upper right:} $5.8\,\mu$m image constructed by us from \textit{Spitzer} data obtained from
the public archive.
\textbf{Lower left:} \textit{Herschel} $70\,\mu$m map.
\textbf{Lower right:} \textit{Herschel} $500\,\mu$m map. The \textit{Spitzer} and \textit{Herschel}
images are displayed with a square-root intensity scale.
 }
              \label{opt-mir-fir}%
    \end{figure*}

   \begin{figure*}
   \centering
   \includegraphics[width=18cm]{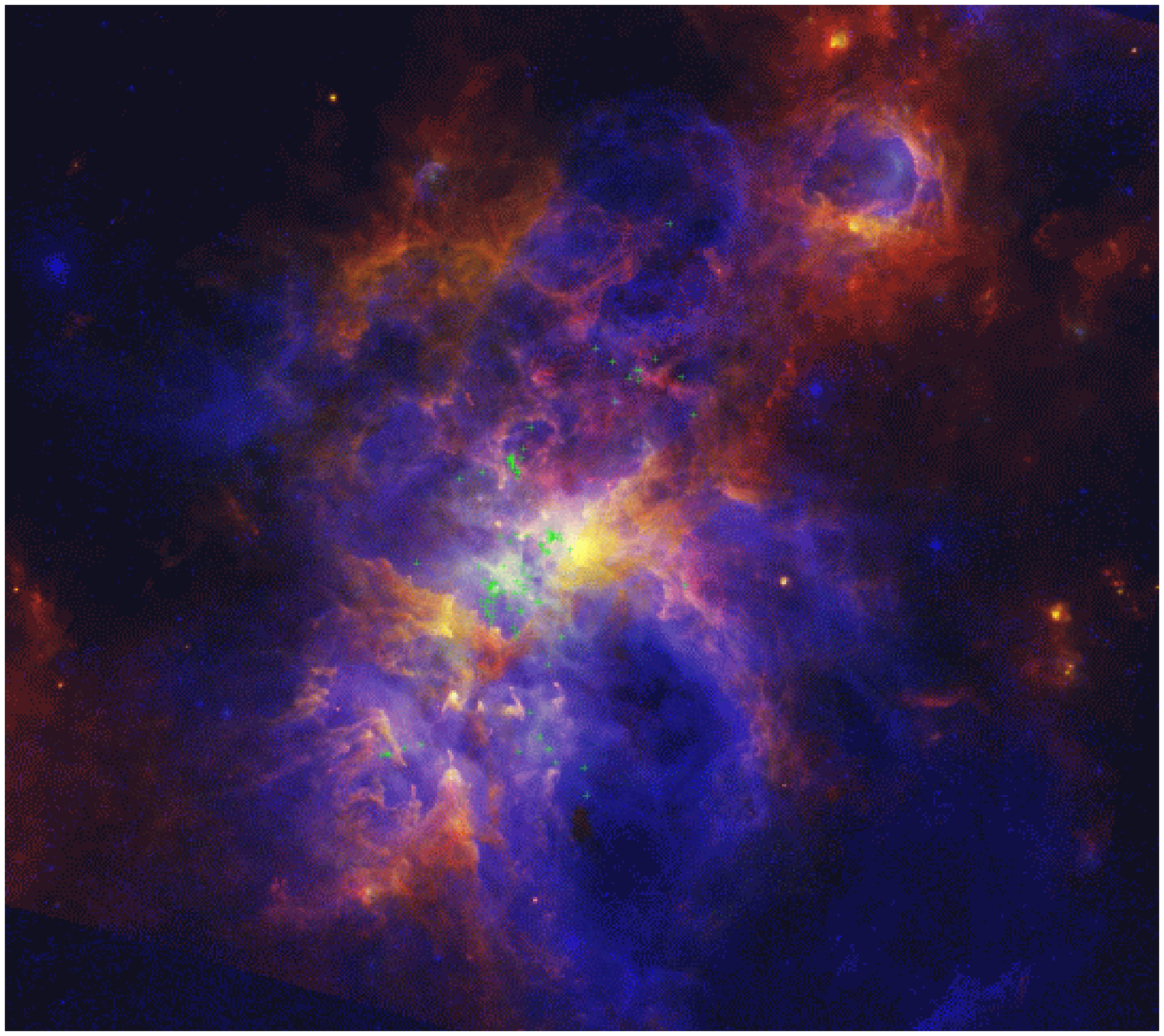}
   \caption{RGB composite of the red optical (DSS) image in blue, the \textit{Herschel} $70\,\mu$m image 
 in green, and the \textit{Herschel} $160\,\mu$m image in red. The image shows a $2.9\degr \times 2.6\degr$
field of view. North is up and east to the left.
 The green plus signs mark the positions of the high-mass stellar
members of the CNC as listed in \citet{Smith06}.
    }
              \label{dss-70-160}%
    \end{figure*}

\section{Cloud morphology}

The complete set of all five \textit{Herschel} maps resulting from our observation
is shown in Figures~A1 to A5 in the appendix.
In Figure~\ref{opt-mir-fir} we compare the $70\,\mu$m and $500\,\mu$m
\textit{Herschel} maps to an optical image and the \textit{Spitzer}
$5.8\,\mu$m image.
The remarkable similarity between the \textit{Herschel} 70 $\mu$m image
and the \textit{Spitzer} $5.8\,\mu$m image demonstrates that
 most of the clouds
are only moderately dense, i.e., are already transparent in the \textit{Spitzer} $5.8\,\mu$m band.
This point will be discussed in more detail in Sect.~5.

In Fig.~\ref{dss-70-160} we show an RGB color composite of the 
optical,  $70\,\mu$m and $160\,\mu$m emission.
This illustrates the spatial relation between the hot ($\sim 10^4$~K) H$\alpha$ emitting
gas and the dense cool/cold gas. It shows how the hot gas fills the interior of the
bubbles in the cloud structure.

The comparison of the \textit{Herschel}\, $70\,\mu$m maps 
to our LABOCA $870\,\mu$m map presented in \citet{CNC-Laboca}
shows that the detectable sub-mm emission 
is restricted to a small volume fraction of the cloud shown by \textit{Herschel}. The volume filling factor of
the dense clouds is very low, and this dense material is highly fragmented
and dispersed throughout the area.
\textit{Herschel} reveals the much more widespread diffuse gas at lower densities.

   \begin{figure*} \centering
   \includegraphics[width=16cm]{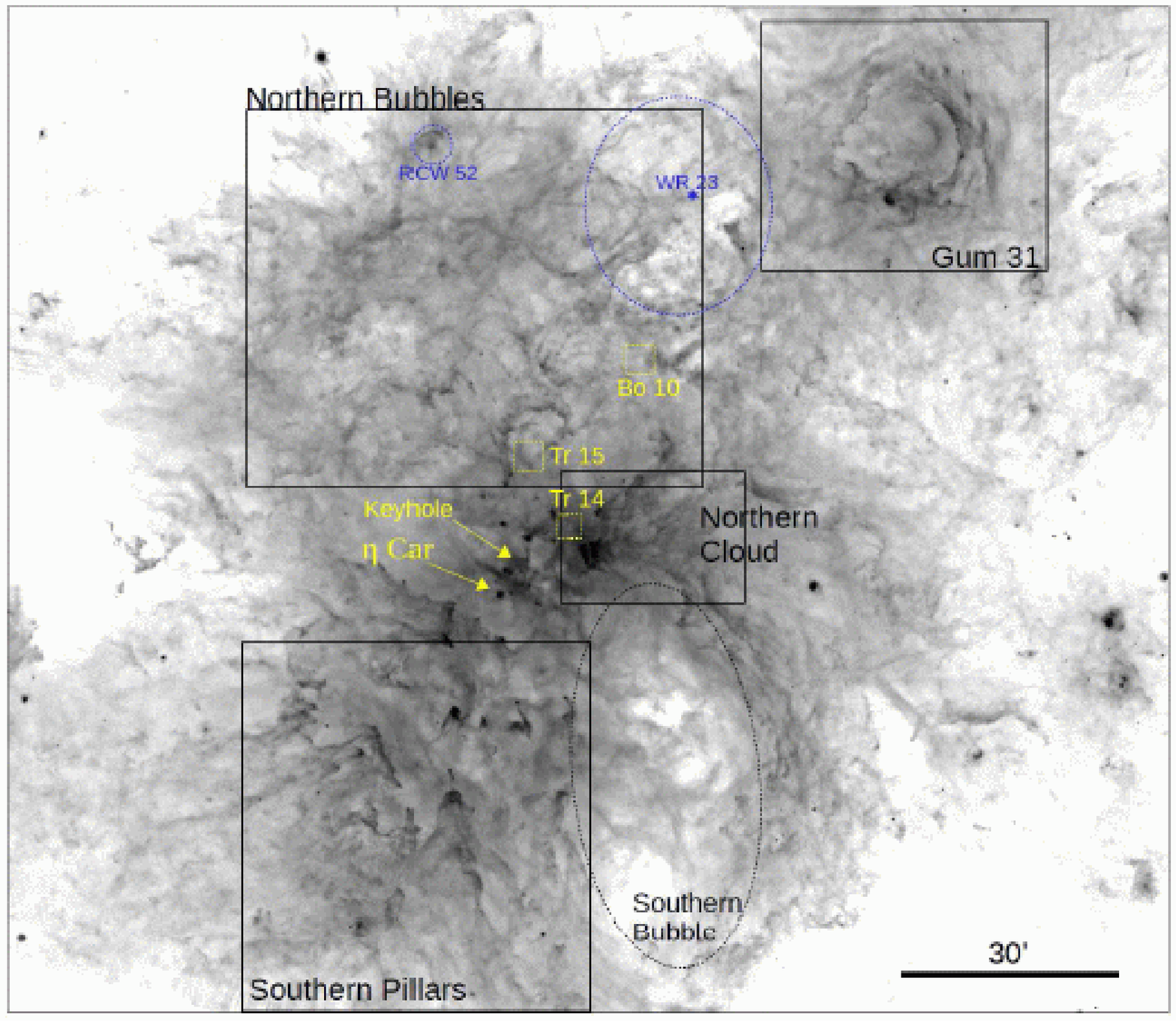}
   \caption{Unsharp-masked version of the \textit{Herschel} $70\,\mu$m image  
    with a logarithmic intensity scale image. The different regions mentioned in
Section 3 are marked by the black boxes. The positions of $\eta$~Car, the clusters Tr~14, Tr~14, and Bo~10,
as well as the HII regions around WR~23 and RCW~52 are marked. North is up and east to the left.
}
              \label{h70-usm}%
    \end{figure*}

\medskip

Figure~\ref{h70-usm} is an unsharp-mask filtered version of the $70\,\mu$m map that
highlights the small-scale structure of the clouds.
It clearly shows that the clouds are highly filamentary and have a very complex morphology.
\textit{Herschel} mapping of other star forming regions has shown that 
molecular clouds generally exhibit an extensive filamentary structure \citep[e.g.][]{Andre10}
and that the
typical width of these filaments is of the order of $\sim 0.1$~pc \citep{Arzoumanian11}.
This implies that the width of the individual filaments can be only marginally resolved in our 
$70\,\mu$m map and remains unresolved in the longer wavelength maps of the CNC.

Based on the observed global cloud  morphology, we define several sub-regions
of the CNC and discuss them separately.

\subsection{Central region}
In optical images of the Carina Nebula, the ``V''-shaped dark
clouds just below $\eta$~Car constitute a very prominent feature. 
The \textit{Herschel} image
shows that these dark clouds are in fact of only moderate density.
As we will find below, their typical visual extinction is just 
$A_V \sim 3 -5$~mag.

The nebulosity to the north of $\eta$~Car is the well known 
``Keyhole Nebula'', which was first described by J.~Herschel in 1840
 \citep[a reproduction of Herschel's drawing and a comparison to a modern
photograph can be found as Fig.~4 in][]{SB08}.
While this nebula contains large amounts of hot gas, producing copious 
H$\alpha$ emission, it also contains cold molecular clouds 
\citep{Cox95,Brooks05,Gomez10,CNC-Laboca}

To the south of $\eta$~Car, the \textit{Herschel} image shows a region
of low emission. This seems to represent a bubble of lower density, that 
was probably created by the wind- and radiation-feedback from the numerous
very massive stars in the Tr~16 cluster.

\subsection{Southern Pillars  (SP)}

The cloud structure in the so-called ``Southern Pillars'' (\textbf{SP}) was already
discussed in detail by \citet{Smith10b} on the basis of the
\textit{Spitzer} maps.
The cloud morphology we see in our \textit{Herschel} maps is very similar.

\subsection{Northern Cloud (NC)}
The cloud to the west of the center, close to the stellar cluster Tr~14, is 
clearly the densest and most massive cloud structure in the complex.
Following \citet{Schneider04}, we denote it as the ``Northern Cloud'' (\textbf{NC}).
The eastern edge, where this cloud faces the stellar clusters, is the site
of a prominent Photon Dominated Region (PDR), which has been studied
in some detail by \citet{Brooks03} and \citet{Kramer08}.
Our \textit{Herschel} maps show that the cloud extends (at least) about
$15'$  ($\approx 10$~pc) to the west.

\subsection{Bubbles north of $\eta$~Car (NB)}

The field north of $\eta$~Car
also shows a few prominent pillars, but the general structure
is dominated by a web of bubbles and arcs. We will therefore denote
this region as the ``Northern Bubbles'' (\textbf{NB}).
The diffuse appearance of the clouds  and the lack of 
sub-mm emission from dense clumps in this region suggests that the
density contrast of the clouds
is smaller than in the SP region, where numerous dense and massive pillars
are present. This morphological differences may indicate
that the massive star feedback in this area
is different from the strong irradiation that shapes the surfaces of the
southern pillars.

The bubble-like structure of the clouds is partly related to individual HII regions
that have swept up shells of dense clouds around them.
The two most prominent HII regions in this area are 
the extended nebula surrounding the Wolf-Rayet star WR~23 (=HD~92809; spectral type WC6),
and the RCW~52 (=Gum~32) nebula (which is illuminated by the O7V star LSS~1887).
Both regions are clearly surrounded by bubbles of gas and dust, which have been
interpreted as being mainly created by the action of the stellar winds of the
massive stars on their environment \citep{Cappa05,Cappa11}.

The irregular filamentary structure could be caused by evolved bubbles
that already broke up into pieces. We note that this region
contains the cluster Tr~15 and Bo~10, which have estimated ages
of $7-8$~Myr \citep{Dias02,HAWKI-survey} and are thus several Myr older
than the clusters Tr~14 and Tr~16 in the center of the Carina Nebula.
This could suggest that at this older age, the stellar-wind feedback        
(especially from
the evolved massive stars) may play an important role.
However, the validity of this interpretation remains unclear, since
recent studies suggest that massive star feedback is usually dominated by ionizing 
radiation \citep[e.g.][]{Martins10} and the
effects of stellar winds are of secondary importance \citep[e.g.][]{Martins12}.

\subsection{The large southern bubble}

In the \textit{Spitzer} images, the large elongated bubble
south-west of $\eta$~Car is a quite prominent hole structure.
Our \textit{Herschel} maps confirm that this structure
is truly an almost empty bubble, and not caused by the shadow of a dark
cloud.
Besides the small dense infrared dark cloud described already in 
\citet{CNC-Laboca}, only very low levels of emission are seen in the inner region
of this bubble in the \textit{Herschel} maps.
The \textit{Herschel} maps suggest that 
the density increases again at the southern edge, i.e.~the bubble wall may be
closed at the southern edge. The diameter along the major axis is about
$50'$, corresponding to a physical length of about 33~pc.

The comparison of our FIR maps to the optical image (Fig.~2) shows that the H$\alpha$ emission
is well aligned with the inner edges of the inner cloud surface; as the
optical emission is strongest at the edges and lower in the central area,
this suggests a shell-like morphology for the hot gas.

The origin of this bubble remains unclear. 
Our search of the SIMBAD database revealed only three known
early-type stars inside this bubble: the O8 star HD~305438,
the B1/2~II star HD~92741, and the B2~III star HD~92877. None of these
seems to be massive enough to have created this large bubble.
One possible explanation is that the hot gas from the winds of the numerous massive
stars in the central area is leaking out and filling this bubble.
An alternative possibility is  that this bubble represents the southern
lobe of a large bipolar HII region that is driven by the 
massive stars in Tr~16.

\subsection{The Gum~31 bubble}

In the north-western part of our \textit{Herschel} maps,
the prominent circular bubble around the HII region Gum~31, which is
produced by the young stellar cluster NGC~3324, is clearly the
dominating cloud structure.
It represents a nice example of a ``perfect bubble''
of dense clouds around an HII region \citep[see, e.g.,][for other examples]{Deharveng09}
that has been swept up by the expansion of the HII region.

It is interesting to note that the \textit{Herschel} maps show elongated 
filamentary clouds that seem to connect this bubble to the central regions of the
Carina Nebula (see Fig.~2).
The idea that the Gum~31 bubble is 
part of the giant Carina Nebula cloud complex is supported by the available
information about the radial velocities of the molecular clouds. According to the
CO maps of \citet{Yonekura05}, the clouds surrounding Gum~31 have radial velocities 
in the range $V_{\rm LSR} = -24 \dots -21\,{\rm km\,s^{-1}}$, which is very similar
to that of the clouds in the central regions of the Carina Nebula, for which a range of
$V_{\rm LSR} = -26 \dots -18\,{\rm km\,s^{-1}}$ is found.
Another interesting aspect is that the densest and most massive
parts of this bubble are found at the south and south-eastern regions of the rim,
i.e.,~just along a line connecting the center of the bubble to the center of the
Carina Nebula. This may indicate some kind of interaction between the clouds
associated with the Carina Nebula and the Gum~31 bubble.
A more comprehensive and detailed investigation of these aspects will be the topic of a
forthcoming paper.


\section{Cloud colors and temperatures}

\subsection{Morphology in multi-color FIR images}

   \begin{figure*}
   \centering
   \includegraphics[width=14cm]{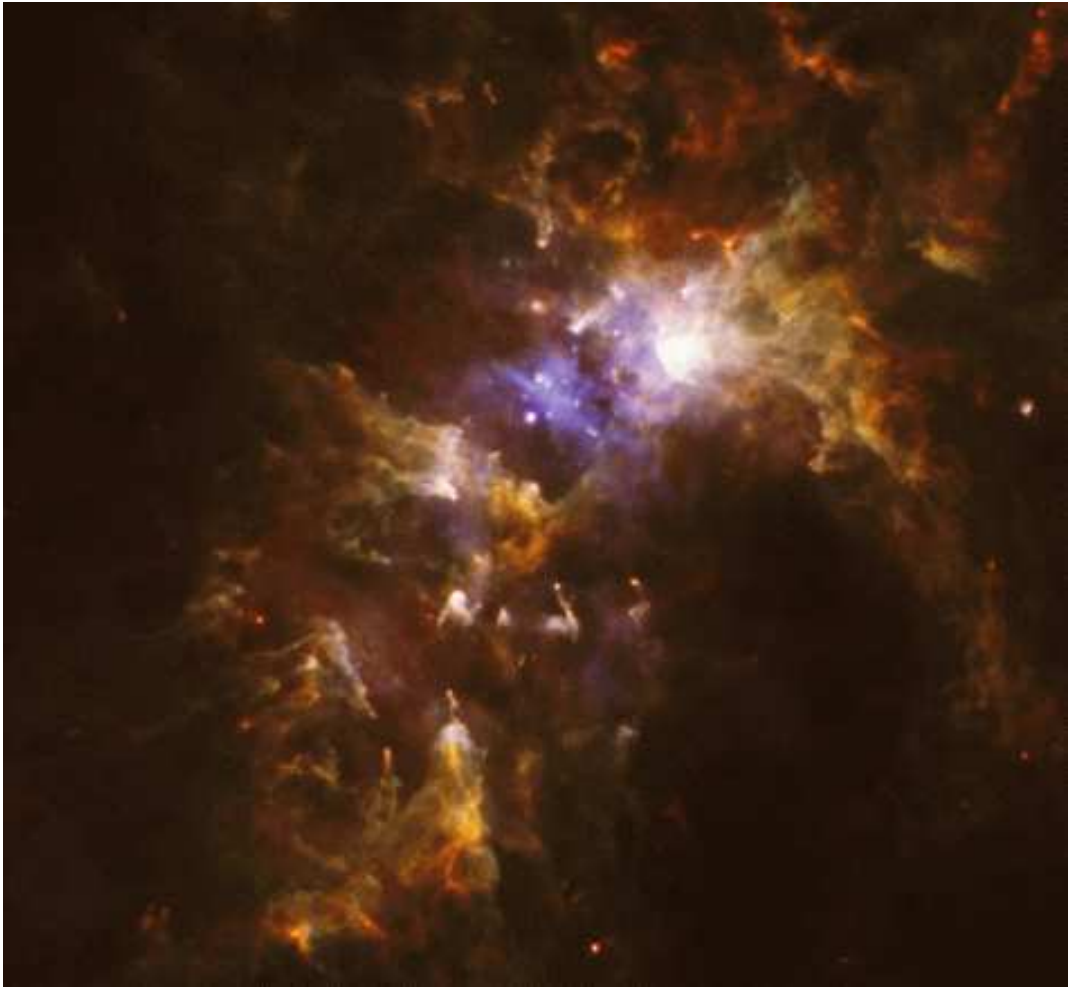}
   \caption{RGB composite of the central region ($1.40\degr \times 1.43\degr$;
north is up and east to the left)
constructed from our \textit{Herschel} $70\,\mu$m (blue),  $160\,\mu$m (green)
and $350\,\mu$m (red).}
              \label{herschel-70-160-350}%
    \end{figure*}

Figure~\ref{herschel-70-160-350} shows an RGB composite of the central region
of the CNC created from our three
shortest \textit{Herschel} wavelengths, in which the colors
trace the different cloud temperatures\footnote{Evaluating the
peak of the Planck function  $B_\nu(T)$ shows that the
  $ 70\,\mu$m band is most sensitive to cloud temperatures of $T \sim 72$~K,
the $160\,\mu$m band to $T \sim 32$~K, the  $250\,\mu$m band to $T \sim 20$~K,
the  $350\,\mu$m band to $T \sim 15$~K, and the $500\,\mu$m band to  $T \sim 10$~K.}.
The bluish emission traces the relatively warm gas in the
Keyhole Nebula Region, and also in front of the eastern edge of the Northern Cloud,
which is strongly irradiated by the stars in the cluster Tr~14.

The red tones in this image trace the coldest cloud material, which is 
concentrated in the densest parts of the base of the pillars in the
Southern Pillar region, and also in a number of compact clouds north of the
Northern cloud.

There are several reddish dense cold clouds to the north-west of center; 
in contrast to similarly dense clouds in the Southern Pillar region, these
do \textit{not} show a pillar-like structure.

\subsection{Color temperatures}

In order to estimate the local temperature of the clouds in our map,
we determined a \textit{color temperature}
from the ratio of the observed fluxes in two \textit{Herschel} bands.
Due to the high level of radiative feedback from massive stars, which
heats the clouds, and the
fact that most clouds have only  moderate column densities, we can
expect that most of the clouds in the CNC should be
\textit{not} extremely cold 
(such as infrared dark clouds 
which often show temperatures of $T \le 15$~K), but somewhat warmer.
 We will therefore employ the
ratio of the $70\,\mu{\rm m}$ versus $160\,\mu{\rm m}$ map intensities, which
is well suited to measure temperatures in the range $\sim 20 - 75$~K\footnote{
For a dust dust emissivity index of $\beta = -2.0$, the $70\,\mu{\rm m}$ versus  $160\,\mu{\rm m}$
flux ratio is 1  for $T = 28.3$~K, 
10 for $T= 74.8$~K, and 0.1 for $T= 18.0$~K.}.
This choice has the advantage that the resulting color temperature maps 
provide the best angular resolution that can be obtained from our \textit{Herschel} maps.

In order to construct a flux-ratio map, we first  convolved 
the $70\,\mu$m map with a kernel as discussed in \cite{Aniano11}
to match the angular resolution of the $160\,\mu$m map.
Following the procedure for the temperature determinations
in the Hi-GAL data described by \citet{Bernard10}, we divided the
fluxes derived from the maps
by the recommended color-correction factors of 1.05 and 1.29 for the 
$70\,\mu$m and $160\,\mu$m PACS bands that were derived from a
comparison of PACS data to \textit{Planck} data.

Assuming that we see optically thin\footnote{The assumption that the clouds are optically thin will
be justified and confirmed in our analysis of the derived column densities in Sect.~\ref{column-densities-sect}}
 thermal dust emission, the ratio 
of the observed fluxes can be related to the temperature via the formula
\begin{equation}
\frac{I_\nu(70\,\mu{\rm m})}{I_\nu(160\,\mu{\rm m})} \propto \frac{B_\nu (T) (70\,\mu{\rm m}) \; \kappa_\nu (70\,\mu{\rm m})}{B_\nu (T) (160\,\mu{\rm m}) \; \kappa_\nu (160\,\mu{\rm m})} = f(T) \label{ctemp.equ} 
   \end{equation}
\noindent
We further assume that the dust emissivity follows a power-law ($\kappa_\nu \propto \nu^\beta$).
The value of the dust emissivity index $\beta$ is not very well known and under debate; typical
observationally determined values range from $\sim 1$ to $\sim 3$, and it is possible
that $\beta$ is anti-correlated to the temperature \citep[e.g.,][]{Shetty09}. Here we assume $\beta = 2$, 
what should be appropriate for the relatively warm clouds in the Carina Nebula
and has been  confirmed  for the not too cold ($T \ge 20$~K) clouds
in HII regions \citep[see, e.g.,][]{Anderson10}.
Equation~\ref{ctemp.equ} allows to determine the temperature in each pixel by comparing the
observed flux ratio with a pre-computed table of flux-ratio values on a finely space temperature grid.

A complication of this temperature determination method results from the
noise in the \textit{Herschel} maps. In regions with low surface brightness, statistical
fluctuations can lead to large errors of the derived temperatures due to division
by very small numbers. Furthermore, some outer regions of the maps contain pixels with negative values, preventing
any temperature determination.
In order to avoid these problems, we restricted the computation of color temperatures
to pixels with intensities above the limits of 
$I_{\nu}(70\,\mu{\rm m}) > 0.001$~Jy/square-arcsec \textit{and}
$I_{\nu}(160\,\mu{\rm m}) > 0.01$~Jy/square-arcsec.
These limits are several times above the noise-level in the maps and also
serve to separate the emission of the clouds in the CNC from the large-scale
galactic background emission.
This minimum intensity condition is fulfilled for 3\,000\,757 pixels in our map, 
i.e.~we can derive color temperatures for a total area of about 2.37~square-degrees
($\approx 44\%$ of the full extent of our maps). Since we are only interested here
in the properties of the relatively bright clouds associated with the Carina Nebula, 
these temperature maps cover nevertheless almost the entire area of interest.
In the much fainter regions, predominantly around the periphery of our map, for which
we cannot determine temperatures, the emission is
dominated by the galactic background, which is not
of interest for our current study.

A potential problem for the color temperatures may be
excess emission from very small dust grains.
The  absorption of individual optical or UV photons
from the surrounding stellar radiation field can cause strong temperature excursions of
the smallest grains (with sizes below about 10~nm).
Due to these stochastic temperature fluctuations by single-photon heating,
the very small grains may not reach an equilibrium temperature.
The emission from very small grains in the high-temperature state, 
immediately after a photon absorption,
can lead to an excess of mid-infrared radiation in the total emission spectrum
of a cloud with a distribution of grain sizes.
As discussed in detail in \citet{Draine07}, two mechanisms play a role.
First, photon-heated very small grains consisting of polycyclic aromatic hydrocarbon (PAH)
can emit strong mid-infrared band radiation, in particular around the wavelengths of
11.3, 12.7, and $17\,\mu$m.
Second, very small carbon and silicate grains 
may cause an excess of mid-infrared continuum emission.
Whereas the \textit{Herschel} bands do not contain strong PAH emission features,
they may be affected by continuum excesses.
The strength of excesses from very small grains depends on the ambient radiation field.
When the rate of starlight heating is large (as is the case in the Carina Nebula), 
the relative amplitude of the temperature fluctuations of the grains decreases and the 
the steady state equilibrium temperature approximation can become valid even for very small grains
\citep[see discussion in Sect.~3 of][]{Draine07}.
According to Table~4 in \citet{Draine07} and in the presence of a radiation field
that is 100 times stronger than in the local solar environment,
a variation of the abundance of very small grains
by a factor of 10 leads to variations in the emissivities of 17\% for the $70\,\mu$m PACS band,
7\% for the $160\,\mu$m PACS band, and $\le 5\%$ for the SPIRE bands.
We note that a 20\% error in the $70\,\mu$m intensity causes a $\sim 5\%$ change in 
the derived color temperature. This represents a lower limit to the accuracy of the
temperature determinations.
\medskip

   \begin{figure*}
   \centering
   \includegraphics[width=14.5cm]{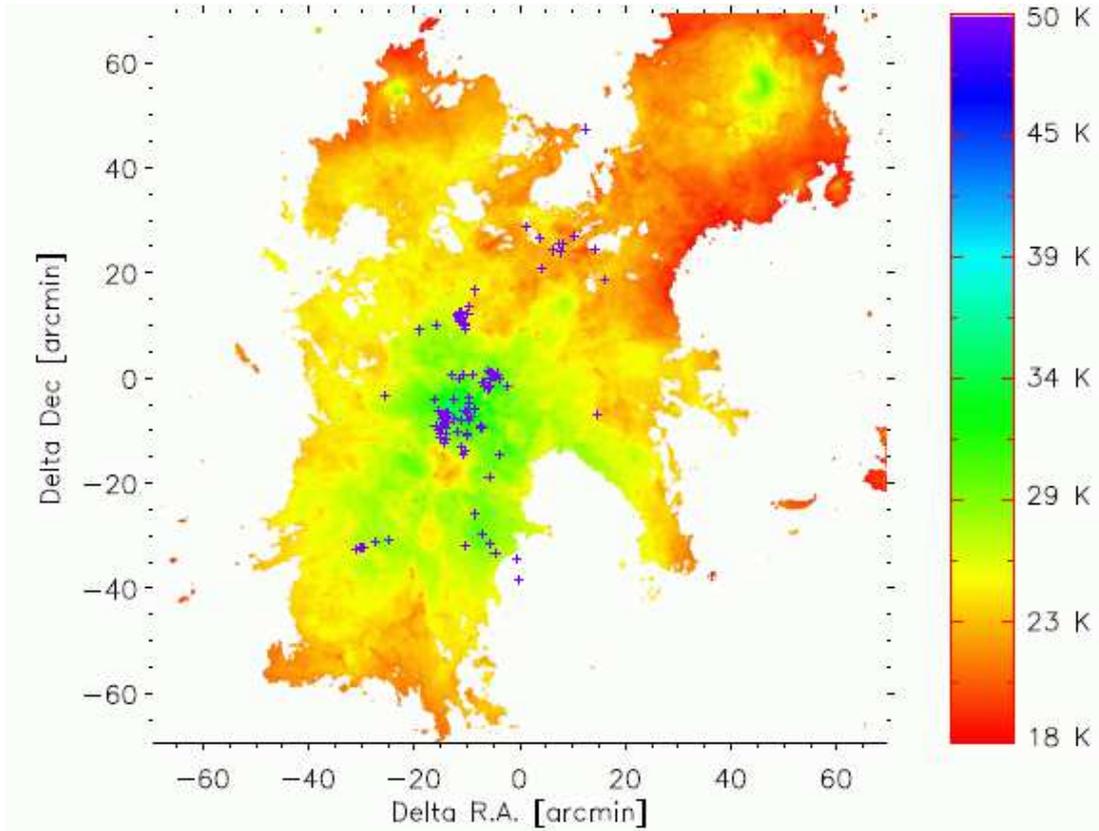}\hspace{3mm}
   \caption{Map of 70--160~$\mu$m color-temperatures. 
Pixels with insufficient $70\,\mu$m or $160\,\mu$m intensities for
a reliable determination of the intensity ratio are left white.
The blue plus signs mark the locations of the high-mass stellar
members of the CNC as listed in \citet{Smith06}. 
    }
              \label{temp-map}%
    \end{figure*}
%
   \begin{figure*}
   \centering
   \includegraphics[width=14.5cm]{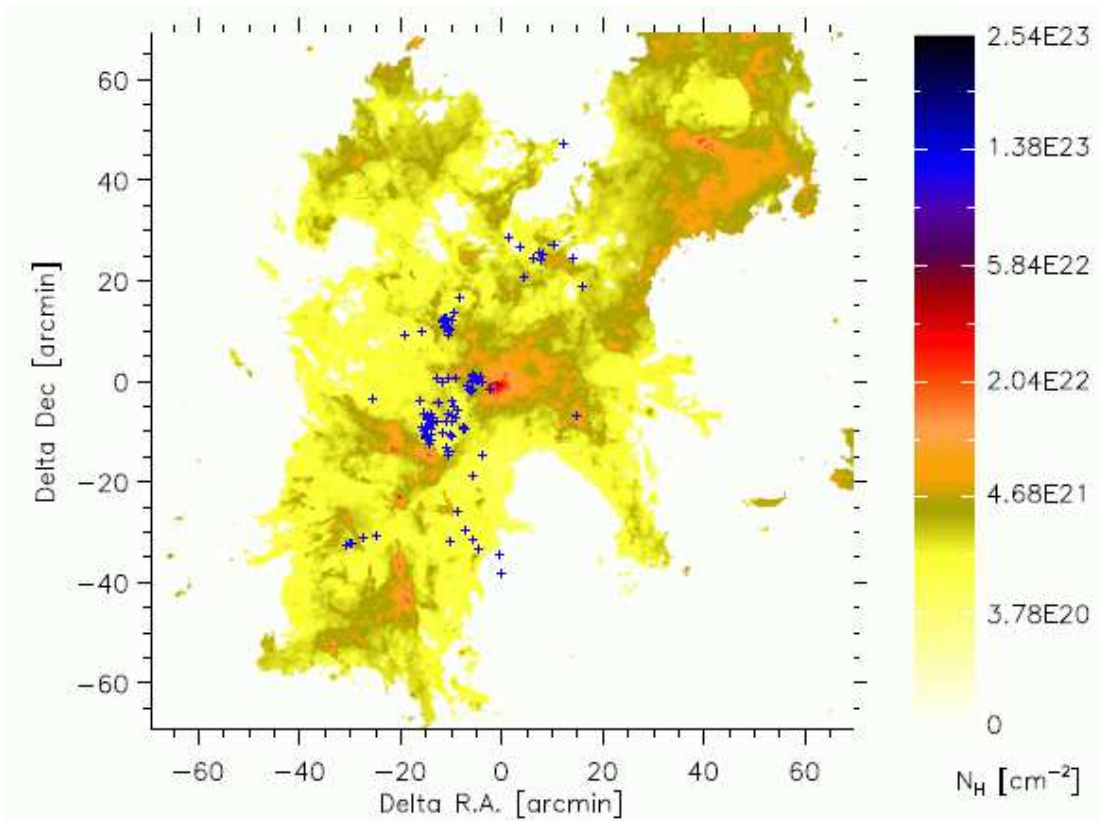}\hspace{3mm}
   \caption{Map of column densities $(N_{\rm H})$
with symbols as in Fig.~\ref{temp-map}.
Pixels with insufficient $70\,\mu$m or $160\,\mu$m intensities for
a reliable determination of the intensity ratio are left white.
    }
              \label{ext-map}%
    \end{figure*}

Figure~\ref{temp-map} shows a map of the resulting color temperature values.
A histogram of the derived color temperatures  can be found in Fig.~\ref{temp-histo}.
The color temperatures range from 14~K to 83~K; the mode and the median of the
distribution is 29~K and 27.3~K, respectively.
Only 4\% of the pixels have $T < 20~K$, and the
fraction of pixels with temperatures of $T > 40~K$  [$50~K$] is just 0.39\% [0.01\%].

In order to estimate the uncertainties of the derived color temperatures, we first note
that the flux calibration of the \textit{Herschel} maps
is thought to be accurate to $\sim 10-20\%$; at
$T \approx 30$~K, a 20\% intensity error causes a 5\%  ($\approx 1.4$~K) change of the
derived color temperature. 
Besides the above mentioned uncertainties related to possible excess emission from 
very small grains, another contribution results from the 
photometric calibration uncertainties of the \textit{Herschel} bands.
For temperatures below $\sim 20$~K, the quickly rising photometric color correction terms
cause higher uncertainties and a possible bias towards over-estimated temperatures;
however, since only 4\% of the pixels in our maps yield such low temperatures,
this should not be a serious problem in the analysis of the global properties of the clouds
presented in this paper.
Furthermore, since our color temperatures are derived from the \textit{Herschel} maps
with the shortest wavelengths, they may be biased towards higher dust temperatures
(and consequently lower column densities). Preliminary results from our more detailed investigation
(that will be presented in a subsequent publication) suggest that any such bias is rather small.
In summary, we estimate the uncertainties of the derived temperatures to
be about $\sim 10 - 15\%$.

We also would like to point out that the temperature values we determined
represent averages over the line-of-sight through the full depth
of the clouds.
Since in some locations there may be several individual clouds that are seen projected onto each other,
the physical meaning of the derived temperature is limited, but this is 
a general problem for the study of cloud properties, and not specific to our data set.
\medskip

Our temperature map in (Fig.~\ref{temp-map}) shows a clear systematic temperature gradient
from the central regions (where most of the hot, massive stars are located)
to the periphery of the complex. A similar, but much steeper temperature gradient is seen within the
boundaries of the Gum~31 nebula. This reflects the fact that the number of high-mass stars in
NGC~3324 (about 3) is much smaller than the 126 O-, early B-, and WR stars \citep{Smith06} 
in the central regions
of the CNC.

\section{Cloud column densities \label{column-densities-sect}}

With the temperature estimate we can now proceed and
compute the column densities of the clouds.
The observed surface brightness in our maps can be converted to
the beam-averaged hydrogen column density via the formula
\begin{equation} 
N_{\rm H} = 2\,N_{\rm H_2} = 2 \, \frac{F_{\nu}\,R}{B_{\nu}(T_d)\;\Omega\;\kappa_{\nu}\;\mu\,m_{\rm H}}\;\; ,
\end{equation}
where $\Omega$ is the beam solid angle, $R$ is the gas-to-dust mass ration
(assumed here to be 100), $\kappa_{\nu}$ is the dust opacity, and
$\mu$ is the mean molecular weight.
For consistency with our previous analysis of the LABOCA sub-mm data
\citep{CNC-Laboca}, we use
the \citet{Ossenkopf94} dust model for grains with thin ice-mantles
that coagulated at a density of $n = 10^6\,{\rm cm}^{-3}$ which gives
a dust opacity of
$ \kappa_{\nu}(160\,\mu\rm m) = 40.5\, cm^2/g$.
We note that, due to the unknown details of the chemical composition
and physical structure of the dust grains, opacity values suffer from
uncertainties of (at least) a factor of $\sim 2$, resulting from the dependence
of the dust opacity on the details of the grain properties.
This causes uncertainties by  a factor of $\sim 2$ for the derived cloud masses.

\begin{figure*}
   \centering
   \includegraphics[width=8.5cm]{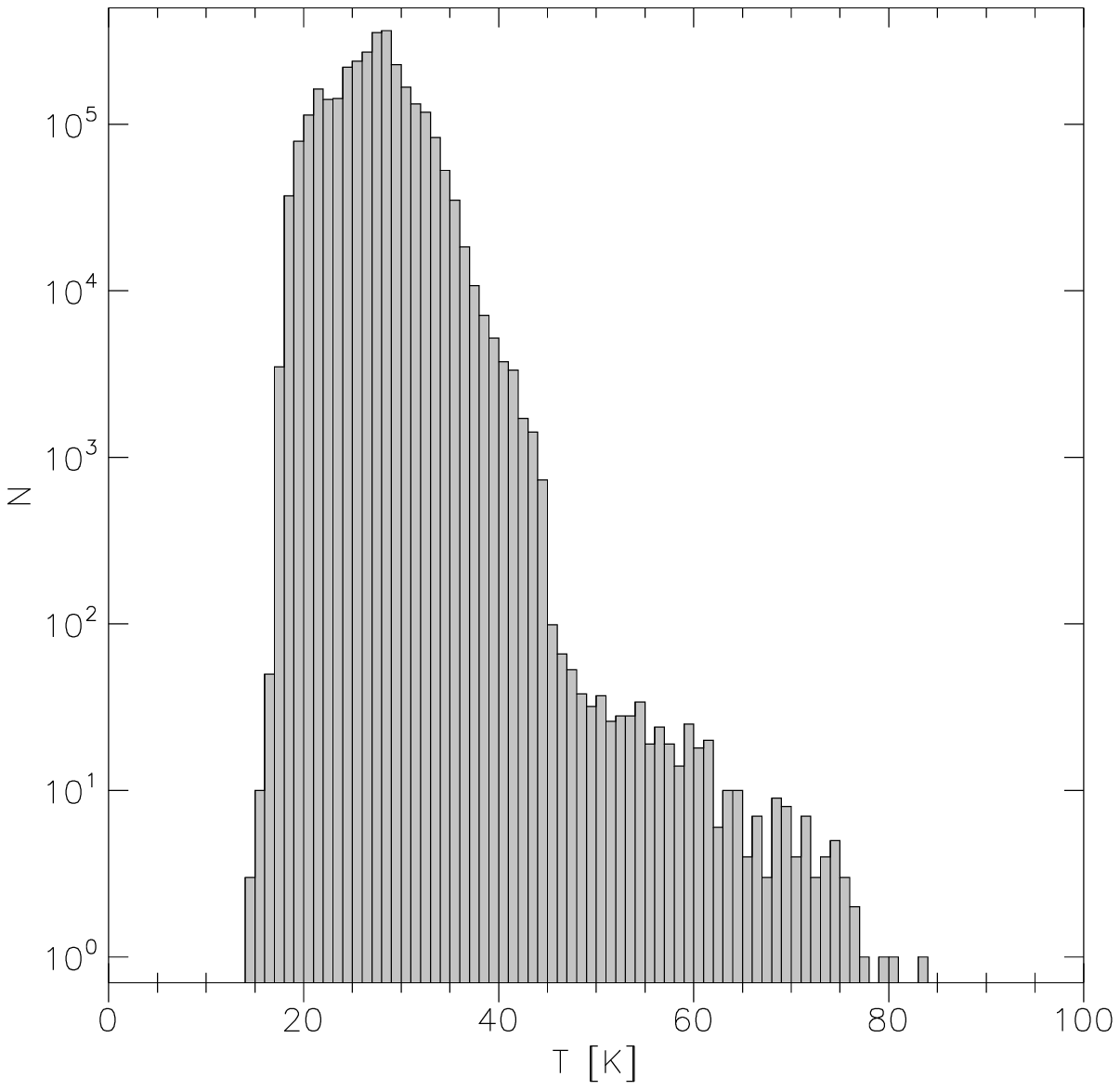}\hspace{3mm}
      \includegraphics[width=8.5cm]{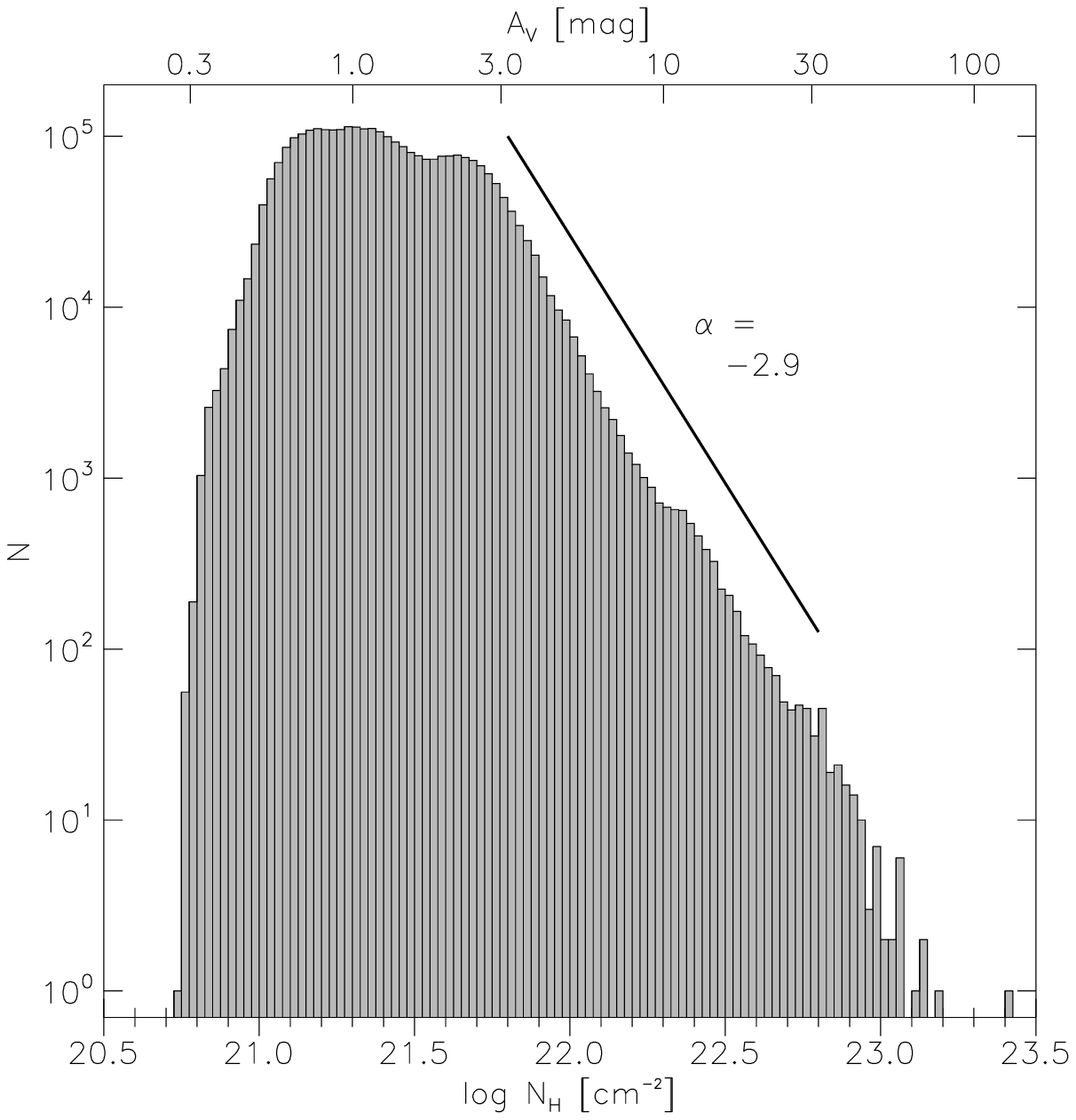}
   \caption{Histogram of 70--160~$\mu$m color temperature values (left) and 
column density values (right) derived from our maps.
    }
              \label{temp-histo}%
    \end{figure*}

Figure~\ref{ext-map} shows the resulting map of cloud column densities.
We assume that column density and visual extinction are related via
the ``canonical'' relation
$ A_V = 1\,\rm mag \, \leftrightarrow \, N_{\rm H} = 2 \times 10^{21}\,{\rm cm}^{-2} \,
\leftrightarrow \, N_{\rm H_2} = 1 \times 10^{21}\,{\rm cm}^{-2}$
\citep[see, e.g.,][]{Bohlin78}.
For the analysis of the statistics of the column densities, we excluded
a $1' \times 1'$ region centered on $\eta$~Car, which is a source of strong
FIR emission.
Fig.~\ref{temp-histo} shows the histogram of the column density values.
We find the mode of the distribution at $A_V = 1.0$~mag and a median value of $A_V = 1.17$~mag.
The distribution shows a second peak near $A_V \approx 2$~mag, and 
drops steeply towards higher column densities.
In the 
$A_V \approx 3 - 30$~mag range, this drops can be well described by 
a power-law relation of the form
\begin{equation}
\frac{d\,\log(N)}{d\,\log(N_{\rm H})} \propto \left( \log(N_{\rm H})\right) ^{\alpha}
\end{equation}
with an exponent
$\alpha \approx -2.9$.
This slope is considerably steeper than column density slopes derived
by \citet{Hill11} from \textit{Herschel} observations of the Vela~C cloud complex 
($\alpha \approx -2.0$)
or \citet{Kainulainen11} from an extinction map of the Ophiuchus cloud ($\alpha \approx -1.5$).
Although a direct comparison of different star forming regions is difficult
due to the unavoidable differences in the observations and data analysis, this result
suggests that the CNC contains a smaller fraction of the total cloud mass
at high densities than the other mentioned cloud complexes. 
This may be the consequence of the fact that the CNC is already a somewhat evolved
star forming region (star formation has started already about 8~Myr ago), much of
the original dense cloud mass has already been dispersed by the very high level of 
massive star feedback, and the today present dense clouds are less the remnants of the
original cloud but more the result of stellar feedback concentrating and compressing
some fraction of the remaining clouds.

Our finding that just
1\% of the pixels have $A_V \ge 5.26$~mag, and only
0.1\% have  $A_V \ge 12$~mag 
agrees with our estimate of the (generally moderate) cloud extinction  mentioned
above and can also be used to provide an (a-posteriori) justification of the
assumption of optically thin FIR emission.
The $70\,\mu$m opacity of the grains in the above mentioned \citet{Ossenkopf94} dust model shows 
that the optical depth of 99.9\% of the pixels in our maps (which have  $A_V < 12$~mag)
is $\tau(70\,\mu{\rm m}) < 0.036$.

   \begin{figure*}
   \centering
   \includegraphics[width=8.5cm]{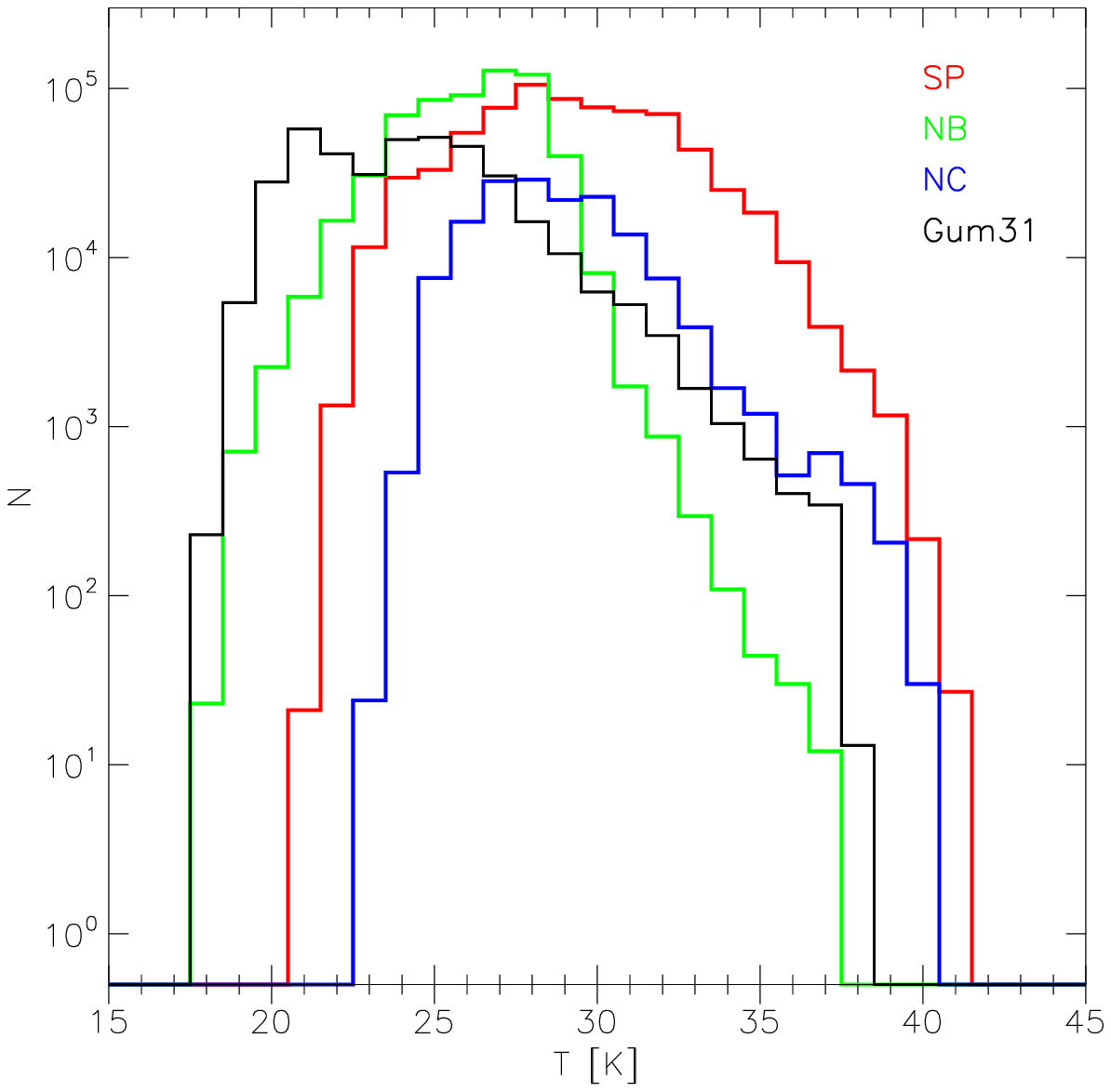}\hspace{3mm}
      \includegraphics[width=8.5cm]{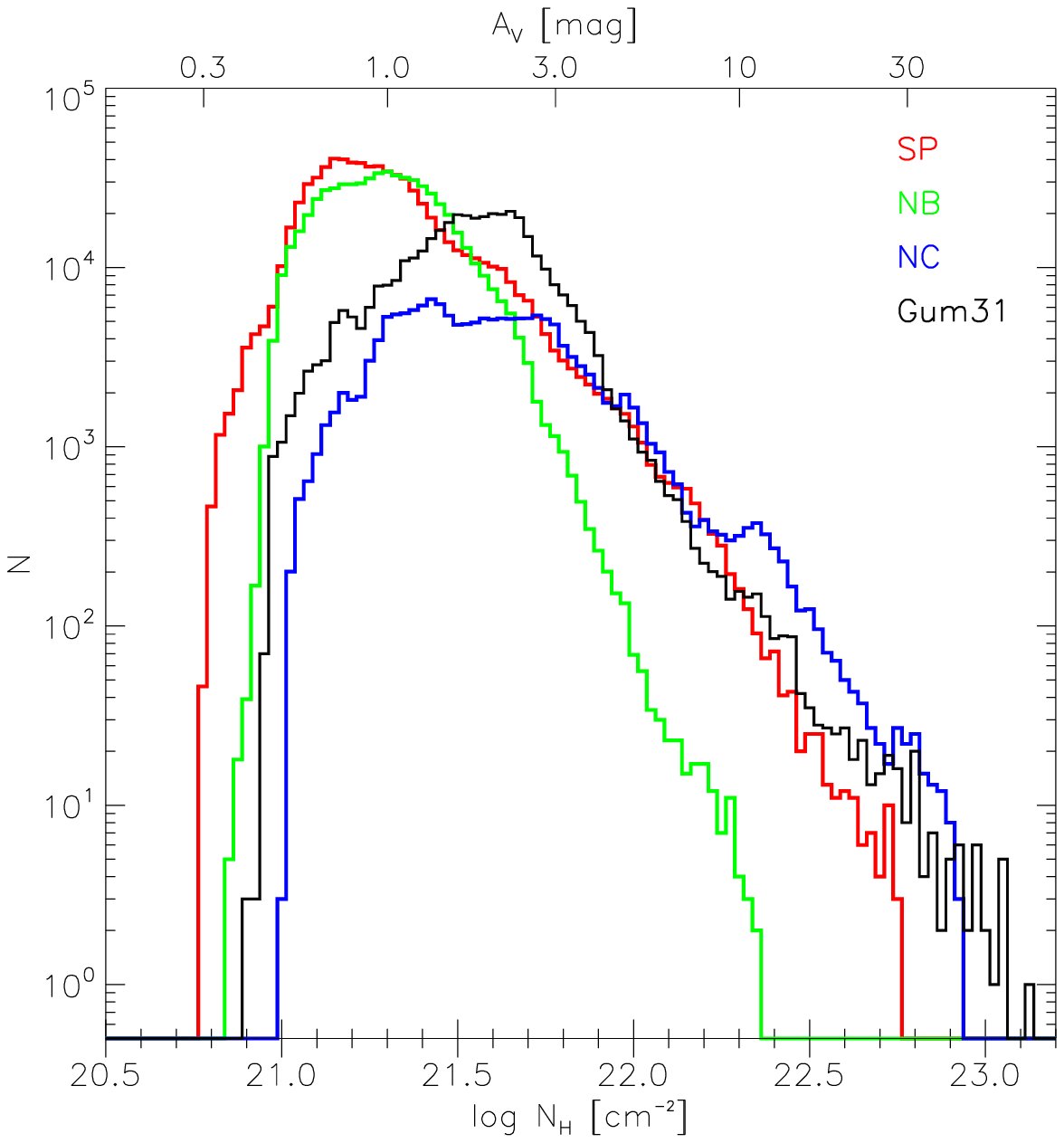}
   \caption{Histogram of 70--160~$\mu$m color temperature (left) column density values (right)
for the individual subregions defined in Sect.~3.
    }
              \label{ext-map-histo-reg}%
    \end{figure*}

\section{Column-density-  and temperature-distributions
and integrated cloud masses
 in the different parts of the complex}

\begin{table}
\caption{Integrated cloud masses for different extinction thresholds.}
\label{tab:cloudmasses}
\centering
\begin{tabular}{l r  r r r}
\hline\hline
\noalign{\smallskip}
Region & {$M_{\rm tot}$}&{$M_{\rm tot}$}&{$M_{\rm tot}$}\\
 & &{$(A_V > 3)$}&{$(A_V > 7)$}\\
 & {$[M_\odot]$}&{$[M_\odot]$}&{$[M_\odot]$}\\
\noalign{\smallskip}
\hline
\noalign{\smallskip}
{CNC ($R= 1\degr$)}  &655\,700&105\,100&22\,600\\       
 {SP}               &209\,900&32\,300&6940\\
 {NC}               &92\,800&44\,500&14\,300\\
 {NB}               &161\,500&4075&185\\
 {Gum\,31}          &186\,700&46\,400&7325\\
\hline                                   
\end{tabular}
\end{table}

We can now compare the properties of the cloud structure in the different parts
of the CNC as defined in Sect.~3.
First, we determine the total cloud masses in each region by integrating our
column densities over the corresponding area.
For the total cloud complex associated to the Carina Nebula we define a
circular region with a radius
of 1 degree, centered on the position R.A.=10h44m16s, DEC=-59d38m28s
(this point is $6.5'$ north-east of $\eta$~Car). This region includes nearly
all the clouds in our map, but excludes the clouds around Gum~31.
Integrating the derived column densities over this circle 
yields a total cloud mass of $M_{\rm tot} = 655\,700\,M_\odot$.
For the Southern Pillars region we find 
$209\,900\,M_\odot$,
and for the Northern Cloud 
  $ 92\,800\,M_\odot$.
%
The integrated mass in the Northern Bubbles region is  $ 161\,500\,M_\odot$,
and for the Gum~31 region we derive 
$186\,700\,M_\odot$.

While the derived integrated masses should be largely free from background contamination
(due to the flux thresholds we applied in our analysis), the gas shows a wide range of
densities, from very diffuse and extended low-density filaments to quite massive clouds
(like the head of the Northern Cloud).
Numerous recent investigations of the relation between cloud structure and the star formation
process lead to suggestions of a (column)-density threshold clouds have to exceed
in order to allow active star formation.
\citet{Lada10} found that the star formation rate in molecular clouds is linearly proportional 
to the cloud mass above an extinction threshold
of $A_V \approx 7$~mag (corresponding to a gas surface density threshold of $\ga 100\,M_\odot\,{\rm pc}^{-2}$). 
The study of \citet{Kainulainen11} suggested the
transition between diffuse clouds and bound clouds (which may be the site of
stars formation in the near future) near $A_V \approx 3$~mag.
We have therefore determined the integrated masses in the above defined regions
for all pixels exceeding the column density thresholds of $A_V = 3$~mag and
 $A_V = 7$~mag. These values are listed in Tab.~\ref{tab:cloudmasses}.

For the full CNC region, the fraction of mass above $A_V = 3 \,[7]$~mag is 16\% [3.4\%].
While these fractions are very similar for the Southern Pillars and just slightly 
higher for the Gum~31 clouds, they are considerably higher (48\% [15\%]) for the
Northern Cloud and considerably lower for the Northern Bubbles (2.5\% [0.1\%]).

In Fig.~\ref{ext-map-histo-reg} we compare the distributions of pixel temperatures and column densities
for the four different regions.
This shows considerable and  remarkable differences between these regions.
The clouds in the Southern Pillars show relatively high cloud temperatures.
We interprete this as a signature
of the strong irradiation that heats these clouds. 
In combination with the relatively low column densities this 
suggests that a large fraction of this cloud material is currently in the process of
being evaporated due to this heating.

The Northern Bubbles are characterized by 
systematically lower cloud temperatures. The distribution of column densities
is quite similar to that in the  Southern Pillars for low extinction values
$(A_V \la 2$~mag), but drops off very quickly for higher values. This implies
that the density contrast in this region is considerably smaller  than in the
Southern Pillars.  This is probably a consequence of the quite different levels and
nature of massive star feedback imposed onto these clouds. In the Southern Pillars,
the clouds are directly irradiated, what heats their surfaces and can lead to
considerable cloud compression in some locations. The Northern Bubbles, on the other hand, are 
apparently in a situation of less strong irradiation and their structure seems to 
be dominated by matter flows.

The temperature distribution in the Northern Cloud is quite similar to that
in the Southern Pillars. The same is true for the distribution of column densities
for values up to $\log \left( N_{\rm H}  \right) \le 22.3$. For higher densities, however,
the Northern Cloud
shows a prominent excess peak near $\log \left( N_{\rm H}  \right) \le 22.4$, which
can be associated to the dense massive center of this cloud.

Finally, the clouds around Gum~31 show relatively low temperatures
(with a mode value of 21~K), and intermediate column densities.

\smallskip

It is also interesting to compare the masses derived here from our
\textit{Herschel} maps to those determined from our previous LABOCA map.
The total mass of all clouds for which our LABOCA map  
of the central $1.25\hbox{$^\circ$} \times 1.25\hbox{$^\circ$}$ observed region
was sensitive
is $60\,000\,M_\odot$ \citep{CNC-Laboca}.
Considering the same area in our \textit{Herschel} maps,
we find here a total mass of $427\,000\,M_\odot$, with
$90\,200\,M_\odot$ above the $A_V \ge 3$~mag threshold.
This agrees well with the expectation that LABOCA traces only the
mass of the denser clouds.

\section{Total cloud mass derived from a simple radiative transfer modeling}

If we want to determine the total mass of all the clouds in the CNC,
we have to take into account that \textit{Herschel} is not very sensitive
to diffuse warm gas at temperatures $\ga 80$~K. However, due to the
very high level of massive star feedback, a substantial fraction of the total
gas mass could be at such relatively high temperatures. This gas should
then be in atomic rather than molecular form.
In order to estimate the total mass of the clouds, we therefore have to
also consider the emission at shorter wavelengths, in the mid-infrared range.

We constructed the spectral energy distribution (SED) of the above described
$1\degr$ radius region (which includes the entire CNC, 
but excludes the clouds around Gum~31) in the following way:
First, we extracted the  total fluxes in this region from our
\textit{Herschel} maps and found values of
844\,500~Jy in the $70\,\mu$m band,
847\,200~Jy in the $160\,\mu$m band,
374\,700~Jy in the $250\,\mu$m band,
156\,000~Jy in the $350\,\mu$m band, and
56\,000~Jy in the $500\,\mu$m band.
For fluxes at shorter wavelengths, we use the
MSX and IRAS fluxes in the $8 - 100\,\mu$m range listed in Table~1 in \citet{SB07}, which were
derived for a very similar area\footnote{\citet{SB07} used a rectangular box aligned
to the galactic coordinate system with
a width of $1.91\degr$  (galactic longitude) and a height of
$2.74\degr$ (galactic latitude), that
includes almost all clouds associated with the Carina Nebula, but excludes 
most of the emission around Gum~31.}.
Finally, we also added the FIR and mm-band fluxes determined in the recent study
by \citet{Salatino12}, who also considered the same area as \citet{SB07}.
The resulting global SED of the CNC is shown in Fig.~\ref{sed.fig}.

In order to estimate the total cloud mass from this SED,
we repeated the radiative transfer modeling as described in \citet{CNC-Laboca}. 
We emphasize that
this radiative transfer modeling 
is not intended to be detailed and highly accurate, but just to
see whether we can reproduce the general shape of the observed
SED with reasonable assumptions about the mass and large-scale
density distribution of the surrounding clouds.
We do not intend to model the small-scale structure of the individual
clouds; instead, we simply assume a central source of radiation
surrounded by a spherical envelope of dust and gas.
Although this is obviously a strong simplification, it provides
the advantage that the temperature distribution
of the gas is computed in a more physically meaningful way
than adding up a few discrete graybody components.

Since a full description of the radiative transfer modeling was already
given in \citet{CNC-Laboca}, we summarize here just the main parameters 
and results:

   \begin{figure}
   \includegraphics[width=8.5cm]{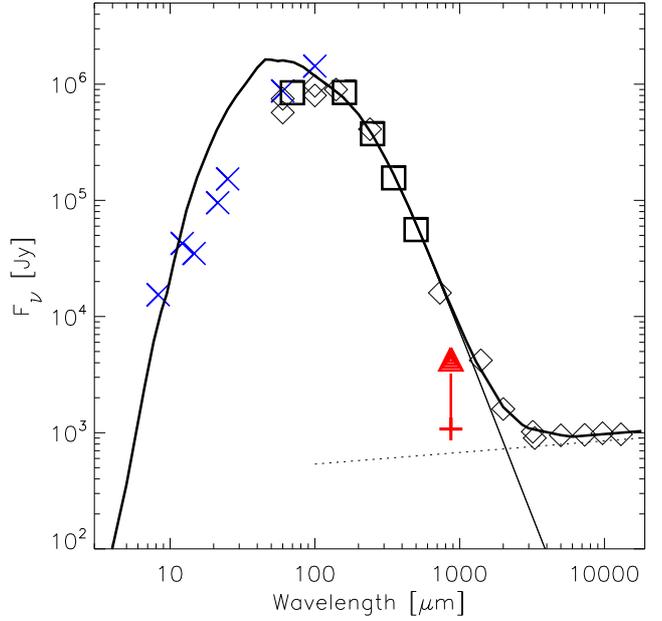}
   \caption{Spectral energy distribution of the Carina Nebula complex.
The thick squares show the fluxes derived from our \textit{Herschel} maps for a
$1\degr$ radius area.
The crosses show the mid- and far-infrared fluxes
determined from the IRAS and MSX maps by \citet{SB07} for a very similar region.
The cross with the
upward triangle shows the lower limit to the total $870\,\mu$m flux derived from our LABOCA
map that covered the central $1.25\degr \times 1.25\degr$ area.
The diamond symbols show the fluxes taken from \citet{Salatino12}. 
The thick solid line is the sum of the
spectrum resulting from the radiative transfer model and
a power-law spectrum of the form  
$F_\nu = 1100~{\rm Jy}\, \left(\frac{3.4\,{\rm cm}}{\lambda}\right)^{-0.1}$
\citep[following][]{Salatino12}.
    }
              \label{sed.fig}%
    \end{figure}

The inner and outer
edges of the spatial grid were set at $r = 10^{18}$~cm (0.3~pc)
and 41~pc.
Following the census of massive stars in the CNC by \cite{Smith06},
we assumed a total stellar luminosity of $2.4 \times 10^7\,L_\odot$
and a typical temperature of $T_{\rm eff} = 44\,700$~K.
The model shown in Fig.~\ref{sed.fig} assumes that the
density increases with distance from the center
according to $\rho(r) \propto r^{1.35}$ for radii up to 20.4~pc and then
stays constant up to the outer radius at 41~pc; this should approximately
match the conditions in the CNC, where most of the cloud material
has already been dispersed from the central region and is now located
at the periphery of the complex.

The model shown in Fig.~\ref{sed.fig} contains a 
total (dust + gas) mass of $889\,140\,M_\odot$.
The spectrum of this simple model
matches the observed FIR fluxes for $\lambda > 100\,\mu$m quite well.
The fluxes at shorter wavelengths are not so well reproduced, but the model fluxes
generally are within a factor of $\sim 2-3$ of the observed values.
We note that a good fit of this part on the SED is not expected from our simple model.
The apparent discrepancy 
is probably the consequence of the fact that the true temperature distribution
is much more complex than our simple model.
Since our model SED predicts systematically too high fluxes in the $15 - 60\,\mu$m range,
the derived mass should be considered to be an upper bound to the total mass.

We also note that the LABOCA $870\,\mu$m flux of 729~Jy is clearly just a \textit{lower limit},
because the LABOCA map covers just the central $\sim 50\%$ of the
analyzed region in our \textit{Herschel} map.
This also explains why the mass derived from this modeling is larger
than our previous estimate ($\la 280\,000\,M_\odot$)
based on the smaller area of the LABOCA map.

It is interesting to compare our result to previous mass estimates
of the CNC based on fits of the SED with a few blackbody curves.
The SED modeling of \citet{SB07}, who used a combination of three discrete 
blackbody curves with temperatures of 35~K, 80~K, and 220~K suggested 
a total (dust + gas) mass of $\approx 962\,000\,M_\odot$, slightly higher
but still  well consistent with
our result of $\leq 890\,000\,M_\odot$.
The SED model of \citet{Salatino12}, which was based on a single
temperature component 
with $T = 34.5$~K, yielded a total dust + gas mass
of $\approx 950\,000\,M_\odot$. 
Despite the uncertainties (e.g., due to the dust opacity),  the results of 
these different approaches to measure the total mass agree remarkably well.

\section{Conclusions and summary}

Our \textit{Herschel} maps show the FIR morphology of the clouds 
in the CNC at unprecedented sensitivity and angular resolution.
They reveal the very complex and filamentary structure of the clouds,
which seems to be dominated
  by the radiation and wind feedback from the numerous massive stars.

The total (gas + dust) mass in the CNC traced by \textit{Herschel} is $\approx 650\,000\,M_\odot$,
and our radiative transfer modeling suggests that, including also warmer gas that is
not well traced by \textit{Herschel}, the total mass may be as high as
$\leq 890\,000\,M_\odot$.
Most of this mass resides in clouds with rather low densities. Nevertheless,
there are about $105\,000\,M_\odot$ in clouds above the extinction threshold
of $A_V > 3$ and about $22\,600\,M_\odot$ above $A_V > 7$. The lower of these two extinction values
($A_V > 3$) is generally considered to constitute the threshold above which the gas
is confined in gravitationally bound entities, making it available for future star formation.
The $A_V > 7$ threshold is though to distinguish the gas that is dense enough to
be directly involved in active star formation. We note that
\cite{Lada10} have demonstrated a good correlation
between the gas mass above $A_V > 7$ and the star formation rate, that
seems to be valid over a very wide range of cloud masses.
This relation predicts a star formation rate
of the order $10^{-3}\,M_\odot\,{\rm yr}^{-1}$ for the CNC, which is well consistent with
the integrated mass and age distribution of the young stars that have formed recently
in these clouds \cite[see][]{HAWKI-survey}.

Another important aspect is the ratio between molecular and atomic gas in the complex.
Our \textit{Herschel} maps trace the FIR dust emission and thus
our total (gas + dust) mass estimate of $\approx 650\,000 - 890\,000\,M_\odot$ 
includes both components,
 i.e.~molecular \textit{and} atomic gas.
The amount of the molecular gas can be inferred from molecular line observations.
\citet{Yonekura05} presented a wide-field $^{12}$CO (J=1--0) survey of the CNC
and determined the total gas masses in different sub-regions.
Our 1 degree region we used for our CNC mass estimates includes 
their regions No 1, 2, 3, 7, as well as some parts of their region 4 (see their Table~1 
and Fig.~3).
Adding up their numbers for the total gas mass based on the $^{12}$CO signal
gives $M_{12\rm CO}\approx 200\,000\,M_\odot$. 
This comparison suggests that about 25\% of \textit{all} the gas in the region is in molecular form,
while 75\% is in atomic form. We note that the quoted percentages suffer from considerable
uncertainties: on the one hand, the CO data may seriously under-estimate the total molecular mass,
and on the other hand, uncertainties of the assumed dust opacity and calibration problems
may affect the estimate of the total gas mass.
Nevertheless, the suggested high mass fraction of atomic gas may be interpreted
 as the consequence of the high level of massive star
feedback that irradiates, heats, and disperses the clouds since several
million years.
The feedback has apparently transformed much, presumably most, of the original molecular cloud mass 
into atomic gas.
\medskip

The \textit{Herschel} maps provide much more detailed information than reported in
this paper. 
The \textit{Herschel} fluxes have been used in our recent study
of the infrared detected jet-driving protostars
in the Carina Nebula \citep{Ohlendorf12}.
In forthcoming publications we will report  results about the detailed small-scale
structure of the clouds, the numerous point-like FIR sources (most of which are expected
to be embedded protostars), as well as the relation between the CNC and the
Gum~31 nebula.

\begin{acknowledgements}
We would like to thank the referee and the editor for providing
several suggestions that helped to improve the paper.
The analysis of the \textit{Herschel} data was funded by the German 
Federal Ministry of Economics and Technology
in the framework of the  ``Verbundforschung Astronomie und Astrophysik''
through the DLR grant number 50~OR~1109.  
Additional support came from funds from the Munich
Cluster of Excellence: ``Origin and Structure of the Universe''.
 
The \textit{Herschel} spacecraft was designed, built, tested, and launched under a contract to ESA 
managed by the Herschel/Planck Project team by an industrial consortium under the overall responsibility 
of the prime contractor Thales Alenia Space (Cannes), and including Astrium (Friedrichshafen) responsible 
for the payload module and for system testing at spacecraft level, Thales Alenia Space (Turin) responsible 
for the service module, and Astrium (Toulouse) responsible for the telescope, with in excess of a hundred 
subcontractors.
PACS has been developed by a consortium of institutes led by MPE (Germany) and including UVIE (Austria); KU Leuven, CSL, IMEC (Belgium); CEA, LAM (France); MPIA (Germany); INAF-IFSI/OAA/OAP/OAT, LENS, SISSA (Italy); IAC (Spain). This development has been supported by the funding agencies BMVIT (Austria), ESA-PRODEX (Belgium), CEA/CNES (France), DLR (Germany), ASI/INAF (Italy), and CICYT/MCYT (Spain).
SPIRE has been developed by a consortium of institutes led by Cardiff University (UK) and including Univ. Lethbridge (Canada); NAOC (China); CEA, LAM (France); IFSI, Univ. Padua (Italy); IAC (Spain); Stockholm Observatory (Sweden); Imperial College London, RAL, UCL-MSSL, UKATC, Univ. Sussex (UK); and Caltech, JPL, NHSC, Univ. Colorado (USA). This development has been supported by national funding agencies: CSA (Canada); NAOC (China); CEA, CNES, CNRS (France); ASI (Italy); MCINN (Spain); SNSB (Sweden); STFC (UK); and NASA (USA).

This work is based in part on observations made with the \textit{Spitzer} Space Telescope, 
which is operated by the Jet Propulsion Laboratory, California Institute of Technology under a contract with NASA.
\end{acknowledgements}

\end{document}